\documentclass[twocolumn,aps,prl,showpacs]{revtex4}

\usepackage{epsfig}

\newcommand{\beq}{\begin{equation}}
\newcommand{\eeq}{\end{equation}}
\newcommand{\bea}{\begin{eqnarray}}
\newcommand{\eea}{\end{eqnarray}}

\begin{document}

\title{Temperature dependence of the nodal Fermi velocity in layered cuprates}
\author{Andrey Chubukov$^1$ and Ilya Eremin$^{2}$}
\affiliation{$^1$
Department of Physics, University of Wisconsin,
Madison, WI 53706 \\
$^2$ Max-Planck Institut f\"ur Physik komplexer Systeme, D-01187
Dresden and Institute f\"ur Theoretische und Mathematische Physik,
TU-Braunschweig, 38106 Braunschweig, Germany}
\date{\today}

\begin{abstract}
We explain recently observed linear temperature dependence of the
nodal Fermi velocity $v_F (T)$ in near-optimally doped cuprates. We
argue that it originates from electron-electron interaction, and is
a fundamental property of an arbitrary 2D Fermi liquid.  We consider
a spin-fermion model with the same parameters as in earlier studies,
and show that the $T$ term is about $30\%$ at $300K$, in
agreement with the data. We show that the  sub-leading term in $v_F
(T)$ is a regular (and small) $T^2$ correction. We also show that at
a $2k_F$ quantum-critical point, temperature corrections to the
dispersion are singular.
\end{abstract}

 \maketitle

The origin of strong deviations from the Fermi liquid behavior in
the normal state of the hole-doped cuprates,  the mechanism of
$d-$wave superconductivity, and the nature of the pseudogap phase
remain the subjects of active debate in the condensed-matter
community. Deep inside the pseudogap phase the cuprates are
Mott-Hubbard insulators. Outside the pseudogap phase Angle-Resolved
Photoemission Spectroscopy (ARPES) and other measurements  show a
large, Luttinger Fermi surface, and $\omega^2$ behavior of the
fermionic damping at the lowest energies~\cite{icc,dess_laser},
consistent with the idea that in this range the system is  a Fermi
liquid with strong correlations.

We take a point of view that the crossover from a metal to a Mott
insulator occurs inside the pseudogap phase, while to the right of
the $T^*$ line, the number of carriers is $1-x$, where $x$ is
doping. In the $1-x$ regime, the fermionic self-energy can be
described in conventional terms, as originating from the interaction
with some bosonic degrees of freedom. A boson can be a phonon, or it
can be a collective electronic excitation in  spin or charge
channels. The same interaction is also thought to be primarily
responsible for the pairing instability, which eventually leads to a
superconductivity.

The nature of the pairing boson in the cuprates is the subject of
outgoing debate,  and in recent years several proposals to
distinguish experimentally between phononic and electronic mechnisms
have been discussed~\cite{timusk,dev,tsv,morr,norm,dev2}. One of the
proposals is to look at the temperature dependence of the
 the Fermi velocity $v_F (T) = v_F (T=0) (1 + \delta (T))$,
 taken along the diagonal of the Brillioine Zone (BZ)
 where the $d_{x^2-y^2}-$wave superconducting gap has nodes
(the ``nodal'' velocity)   It has
been measured recently by Plumb et al.~\cite{dess} in
optimally-doped Bi$_2$Sr$_2$CaCu$_2$O$_{8+\delta}$ be means of
laser-based angle-resolved photoemission spectroscopy. They found
that $\delta (T)$ is approximately linear in $T$ up to at least
$300K$. The linear behavior holds down to $T_c$, and the slope is
quite large: $\delta (T)$ is about $0.35$ between T$_c$ and $T
=250K$. The linear $T$ dependence of $\delta (T)$ is a challenge to
theorists, as on general grounds one would expect an analytic, $T^2$
dependence at the lowest $T$. The magnitude of $\delta (T)$ is
another challenge, as the coupling to the boson is set by  fits to
other experimental data, including the value of $T_c$.

The linear in $T$ dependences of various observables in the normal
(non-pseudogap) state have been reported before and
phenomenologically described in terms of  marginal Fermi liquid
(MFL) behavior~\cite{mfl,vdm}. The linear in $T$ behavior of the
velocity is different by two reasons. First, the self-energy which
yields $\delta (T) \propto T$ scales as $Re\Sigma \propto \omega T$,
while MFL behavior originates from $Re \Sigma \propto \omega
\log{{\text max}~(\omega,T)}$ (and  $Im \Sigma \propto {\text
max}~(\omega,T)$). Second, the linear behavior of $\delta (T)$ has
been measured down to energies ($\pi T \rightarrow \omega$) where
$Im \Sigma$ already has a Fermi liquid, $\omega^2$ form~\cite{dess}.
From this perspective, the linear in $T$ dependence of the velocity
appears to be a truly low-temperature asymptotic behavior, as
opposed to MFL behavior, which likely  holds only above the upper
boundary of the Fermi liquid~\cite{acs,norm}.

The quasi-linear $T$ dependence of the Fermi velocity can be
obtained due to electron-phonon iteraction, but only as approximate
behavior in a limited $T$ range above the Debye frequency
$\omega_D$~\cite{phon_T}, where the temperature-dependent velocity
slowly approaches its bare value after passing through a minumum at
$T \sim \omega_D$.  On the other hand, electron-electron interaction
in 2D gives rise to a linear in $T$ dependence of $\delta (T)$ down
to the lowest temperatures, at which the system is still in the
normal state~\cite{cm,dass}. The linear in $T$  correction to the
velocity comes from  $\omega T$ term in the real part of
self-energy, which in turn originates  from the non-analytic $Im
\Sigma (\omega) \sim \omega^2 \log \omega$, coming from
backscattering~\cite{corr1,cm}.

Although the original analysis in~\cite{cm,dass} was a weak-coupling
perturbation theory, one can show that $\delta (T)$  is linear in
$T$ in an arbitrary Fermi liquid. The reasoning parallels the one in
Ref. \onlinecite{cmgg} for the specific heat coefficient, which is
also linear in $T$ in a 2D Fermi liquid.

The magnitude of $\delta T$ is a different issue. To second order in
the interaction $U(q)$, the velocity renormalization is given
by~\cite{cm}
\begin{equation}
\delta (T) = A T~ \left(\left(U(0)- \frac{1}{2} U(2k_F)\right)^2  + 3  \left(\frac{U(2k_F)}{2}\right)^2\right)
\label{n_11}
\end{equation}
where $A = p_F \log 2/(4\pi^2 v^3_F)$. The two  terms in
Eq.(\ref{n_11}) account for the contributions from charge and spin
channels. Approximating the measured FS of
Bi$_2$Sr$_2$CaCu$_2$O$_{8+\delta}$ (Bi2212) by a circle with $p_F
\sim \sqrt{2} \times 0.6 \pi = 2.7$ (Ref. \onlinecite{FS}, we set
interatomic distance to one), $U(q)$ by $U \sim 2eV$, and using
experimental $v_F \sim 1 eV$ (Ref. \onlinecite{vF}), we find $\delta
(T) \sim 1.6 10^{-5} T$, where $T$ is measured in Kelvins. This
yields $\delta (T) \sim 5 \times 10^{-3}$ for $T \sim 300K$, two
orders of magnitude smaller than the experimental value.

The second-order estimate, however, is only valid in the weak
coupling regime, when the system is far from a Pomeranchuk-type
instability. There is a consensus among researchers that near
optimal hole doping, the coupling is actually strong. Strong
interaction brings about two effects. On one hand, larger coupling
leads to a larger $\delta (T)$. On the other hand, the effective,
screened interaction between quasiparticles deviates from the
second-order $U^2 \Pi$ form,  where $\Pi$ is the dynamic
polarization bubble, and one has to consider the full effect of the
screening. Since $\Pi$ is temperature-dependent, this generaly
affects  the functional form of $\delta (T)$. The purpose of this
paper is to analyze the interplay between these two effects.

As a model for the dynamic screening, we consider the interaction
between fermions and their collective bosonic excitations in the
spin channel. Several groups argued\cite{acs,ilya,che} that this
interaction is responsible for the normal state self-energy and the
pairing. We show  that the velocity renormalization is {\it not}
affected by the full dynamic screening of the effective interaction.
The extra terms in $\delta (T)$ due to the difference between the
full interaction and $U^2 \Pi$, scale as $T^2$ and are small for all
experimentally studied $T$. For the linear in $T$ term, we find,
using the same parameters as in earlier studies, $\delta (T) \approx
 0.37$ at $250K$, in a very good agreement with the data.

The singular interaction in the spin channel is a $2k_F$ process,
which for a hole-like cuprate Fermi surface is an umklapp scattering
between nodal fermions at ${\bf k}_F = (k_{F}, k_F)$ and ${\bf Q} -
{\bf k}_F$. The normal state fermionic self-energy due to
nodal-nodal interaction has been previously considered in the
context of quantum-critical (QC) phenomena \cite{ioffe,krotkov}.
Here we consider the system behavior away from the QC point, when
the bosonic propagator has a finite mass. In the notation $G^{-1}
({\bf k}, i\omega_m) = i\omega_m - \epsilon_{\bf {k+ k}_F} + \Sigma
({\bf k}, \omega_m)$, the  self-energy due to  coupling to spin
fluctuations is given by
\begin{eqnarray}
\Sigma ({\bf k}, i \omega_m) & = &  -\frac{3 {\bar g} }{4\pi^2} T
\sum_{\Omega_m}  \int d^2  q \, \chi ({\bf q} -2{\bf
k}_F, i\Omega_m) \times  \nonumber\\
&& \frac{1}{i (\omega_m + \Omega_m) - \varepsilon_{-{\bf k}_F +{\bf
k + q}+{\bf Q}}} \label{1}
\end{eqnarray}
where ${\bf k_F}$ is the Fermi momentum  along zone diagonal,
counted from $(\pi,\pi)$ point [$(0.8\pi, 0.8\pi)$ for optimally
doped Bi2212]. Near the Fermi surface
\begin{eqnarray}
\varepsilon_{{\bf k + k}_{F}} & = & v_F k_x + \beta^2 k^2_y,
\nonumber
\\ \varepsilon_{{\bf -k_F +k + q + {\bf Q}}} &
 = & - v_F (k_x + q_x) + \beta^2 (k_y + q_y)^2
\label{2}
\end{eqnarray}
where $x$ is set along the zone diagonal, towards $(-\pi,-\pi)$, and
$\beta$ parametrizes the curvature of the Fermi line, $\beta^2 =
1/(2m)$ for a circular FS.

Further,  $\chi ({\bf q}, \Omega_m)$, normalized to $\chi (0,0) =1$,
is the dimensionless dynamical spin susceptibility, and  ${\bar g} =
(U(2k_F)/2) *K$ is the effective, enhanced coupling in the spin
channel. The factor $K$, which reduces to 1 at weak coupling, is the
ratio of the actual and bare static spin susceptibilities at
momentum $2k_F$. We assume, based on neutron scattering data,  that
the momentum dependence of the static susceptibility is weak near
${\bf q} =0$, and approximate the full dynamic spin susceptibility
at small $ {\bf q}$ as $\chi ({\bf q}-2{\bf k}_F, i\Omega_m) =
1/\left(1 +  {\bar g} \Pi ({\bf q} -2{\bf k}_F, i\Omega_m)\right)$,
where $\Pi ({\bf q} -2{\bf k}_F, i\Omega_m)$ is the polarization
bubble with momenta near $2{\bf k}_F$ (the spin factor of $2$ is
included into $\Pi$).

The $2k_F$ particle-hole bubble has been calculated
before\cite{ioffe,krotkov}. It contains a regular part, which  plays
no role in our analysis and which we neglect, and a non-analytic
part (a dynamic Kohn anomaly), which at finite $T$ is given by
\begin{eqnarray}
\Pi_{NA} ({\bf q}-2{\bf k}_F, i\Omega_m) = \frac{1}{4\pi v_F \beta}
\int_{-\infty}^{+\infty} d u \frac{\sqrt{\sqrt{u^2 + \Omega^2_m}+
u}}{4 T \cosh^2{\frac{u-E_{\bf q}}{4 T}}} \label{3_1}
\end{eqnarray}
At $T=0$, this reduces to
\begin{eqnarray}
\Pi^{T=0}_{NA} ({\bf q}-2{\bf k}_F, i\Omega_m) = \frac{1}{2\pi v_F \beta} \sqrt{E_{\bf q}
+ \sqrt{\Omega^2_m + E^2_{\bf q}}} \label{3_2}
\end{eqnarray}
where $E_{\bf q} = - v_F q_x + \beta^2 q^2_y/2$ \cite{comm_1}. The
static $\Pi^{T=0}_{NA}({\bf q}-2{\bf k}_F, 0)$ is nonzero at the
smallest $q$  only for $q_x <0$ (a static Kohn anomaly).
%AC
The non-analycity that gives rise to our effect originates from the form of
$\Pi^{T=0}_{NA}$ at $E_{\bf q}<0$, and $\Omega^2 \ll E^2_{\bf q}$. In
 this limit $\Pi^{T=0}_{NA} \propto |\Omega_m|/\sqrt{|E_{\bf q}|}$, which for
$q_x \sim q^2_y$ is of order $ |\Omega_m|/|q_y|$. The existence of
non-analytic $|q_y|$ in the denominator implies that
$\Pi^{T=0}_{NA}$ gives rise to {\it dynamic} long-range interaction
between fermions.

For the velocity renormalization, we need the real part of the
self-energy Re$\Sigma ({\bf k}, \omega)$ on the mass shell, to first
order in fermionic frequency $\omega$:  Re$\Sigma (\varepsilon_k
=\omega, \omega) = \omega \lambda (T)$.  The temperature variation
of the velocity is related to $\lambda (T)$ as
\begin{equation}
\delta (T) = \frac{v_F(T) - v_F (T=0)}{v_F (T=0)} = -  \frac{\lambda
(T) - \lambda(0)}{1 + \lambda (0)}, \label{n_n1}
\end{equation}
where  $v_F (T=0) = v_F/(1 + \lambda (0))$. The zero-temperature
$\lambda (0)$ is cutoff-dependent~\cite{krotkov}. The
temperature dependence of $\lambda$ on the other hand
 comes from  processes near the
Fermi surface, and is not sensitive to a cutoff.

We first obtain   Im$\Sigma ({\bf k}, \omega)$  using spectral
representation, and then obtain Re$\Sigma$ by Kramers-Kronig
transformation. Substituting Eqs. (\ref{2})-(\ref{3_1}) into Eq.
(\ref{1}), using spectral representation, evaluating the frequency
sums, and re-scaling, we get
\begin{widetext}
\begin{eqnarray}
\mbox{Im} \Sigma ({\bf k}, \bar \omega) = -
\frac{6}{\pi}~\frac{T^2}{T_1} \int _{-\infty}^{+\infty} dt \left[n_B
(t) + n_F (t + {\bar \omega})\right] \int_{0}^{L/\sqrt{T}} \frac{dy
\mbox{Im} Z}{\left(1 + \left(\frac{T}{T_1}\right)^{1/2} \mbox{Re}
Z\right)^2 +
\left(\left(\frac{T}{T_1}\right)^{1/2} \mbox{Im} Z\right)^2}\quad, \label{n_1} \\
Z =  \int_{-\infty}^{+\infty} \frac{dz}{cosh^2{z}} \sqrt{4 z + t +
{\bar \omega} +  {\bar k} - y^2/2 + \sqrt{(4 z + t + {\bar
\omega} + {\bar k} - y^2/2)^2 - (t + i\delta)^2}}\quad,
\label{n_2}
\end{eqnarray}
\end{widetext}
where
%AC
$t = \Omega/T$ is the running dimensionless frequency variable, $y
\propto q_y/\sqrt{T}$ is the dimensional momentum variable along the
Fermi surface, $n_{F,B} (t) = (e^t \pm 1)^{-1}$ are Bose and Fermi
functions, respectively. $L$ is the upper limit of the momentum
integral along the Fermi surface, $T_1 = (4 \pi v_F \beta/{\bar
g})^2$, and ${\bar \omega} = \omega/T$, ${\bar k} = v_F k/T$. The
second-order perturbative result is obtained by neglecting $Z$ in
denominator of (\ref{n_1}).

If the integral over $y$ was convergent, the result of
$y-$integration would be $O(t)$. Im$\Sigma$ would then be determined
by  $t = O(1)$ and have a Fermi liquid form, $\omega^2 + (\pi T)^2$.
Re$\Sigma$ would then only contain a regular $\omega T^2$ term, and
the renormalization of the Fermi velocity would be $T^2$. On a more
careful look, however, we find that at large $y$, $Z$ scales as
$-2it/y$, and the $y$ integral is logarithmic.
%AC
As $t/y \propto |Omega|/|q_y|$, this
logarithmic singularity indeed originates from long-range
dynamical interaction given by $\Pi_{NA}$.

 We assume and then
verify that the logarithmic accuracy is sufficient for the
$T$-dependence of the velocity renormalization, and that the
logarithm is cut at the lower end by $|t|$. To this accuracy,
Im$\Sigma ({\bf k}, \bar \omega)$ is independent on $k$, even in
frequency, and the frequency dependence comes only via the Fermi
function in (\ref{n_1}). Evaluating the integral over $y$ in
(\ref{n_1}) with logarithmical accuracy and subtracting regular
$\omega^2 + (\pi T)^2$ terms, we obtain for the non-analytic part of
Im$\Sigma$
\begin{eqnarray}
\lefteqn{\mbox{Im}\Sigma ({\bar \omega})_{NA} = -\frac{3T^2}{\pi
T_1}\int_0^{L^2/T} dt \times }  && \nonumber\\
&& t \log{t^2} \left[n_F (t + {\bar \omega}) + n_F (t - {\bar
\omega}) -2 n_F (t)\right] \label{n_3_1}
\end{eqnarray}
Substituting (\ref{n_3_1}) into Kramers-Kronig (KK) relation
$\mbox{Re} \Sigma ({\bar \omega}) = (2{\bar \omega}/\pi)
\int_0^\infty \mbox{Im} \Sigma (s)/(s^2-{\bar \omega}^2)$ and
integrating over $s$, we obtain after straightforward calculations
that  at small $\omega/T$
\begin{equation}
\mbox{Re} \Sigma (\omega) =  \frac{6 \omega T}{T_1} \int_0^{L^2/T}
\tanh{t} dt =  \left(\frac{6 L^2}{T_1}\right) \omega - \left(\frac{6
\log 2}{T_1}\right)
  \omega T  \label{n_4}
\end{equation}
The first term is a cutoff-dependent zero-temperature contribution
$\lambda (0) \omega$. The second term, on the other hand, is a
universal, cutoff-independent $\omega T$ term which gives rise to a
linear in $T$ correction to the velocity:
\begin{equation}
\delta (T) =  \frac{ 6 \log 2}{1 + \lambda (0)}~\frac{T}{T_1}
\label{n_22}
\end{equation}
We emphasize that this universal term comes from the {\it upper}
limit of the integration over frequency variable $t$, justifying our
assumption that the logarithmic integral over $y$ is cut by $t$
rather than by external parameters ${\bar \omega}$ and $v_F {\bar
k}$.

To estimate the slope, we recall that ${\bar g} = (U(2k_F)/2) *K$,
where $K$ is the ratio of the actual and bare static spin
susceptibilities at momentum $2k_F$. The bare susceptibility $\chi_0
= p_F/\pi v_F \sim 0.9 states/eV$. The measured $\chi$  has a flat
top between $(\pi,\pi)$ and $2k_F$, at about $13
states/eV$~\cite{abanov}. Using the ratio as the estimate for $K$,
and $U(2k_F) \sim 2 eV$, $p_F \sim \sqrt{2} * 0.6 \pi/a = 2.7$, $v_F
\sim 1 eV$ ($E_F \sim 1.35 eV$), we obtain  $T_1 \sim 2.5*10^3K$
(Ref. \onlinecite{comm_numb}). Using next $\lambda (0) =0.7$
extracted from ARPES fits~\cite{kordyuk}, we find  $\delta T \approx
0.15 \times 10^{-2} T$. For $T =250K$, this gives $\delta (T)
\approx 0.37$, in a good agreement with the data.

We next consider the subleading terms. By  power counting, the
subleading terms, obtained by expanding in $Z$ from the denominator
in Eq.(\ref{n_1}), should scale as  $\sqrt{T/T_1}$.  As $T_1 \sim
2500K$ and the measurements are performed up to $T \sim 300K$, such
corrections would be substantial. We argue, however, that the
corrections to Eq.(\ref{n_22}) are in fact regular $T^2$ terms.  To
see this, we observe that both Re$Z$ and Im$Z$ in Eq. (\ref{n_1})
vanish at large $y$, such that expanding the denominator in
(\ref{n_1}) in powers of $Z$ and integrating over $y$, we loose the
$\log t^2$ term, which was the source of non-analyticity.
Integrating further over $t$ we find that the expansion of Im$
\Sigma (\omega)$ is regular and  holds in even powers of $\omega^2$
and $T^2$, in which case the renormalization of the Fermi velocity
holds in powers of $T^2$.  Alternatively speaking, the $T$ term in
$v_F (T)$ comes from $Im Z$ term in Eq.(\ref{n_1}), while
$Z$-dependent terms in the denominator in (\ref{n_1}) only give rise
to regular, $T^2$ corrections to the Fermi velocity.

\begin{figure}[t]
\epsfig{file=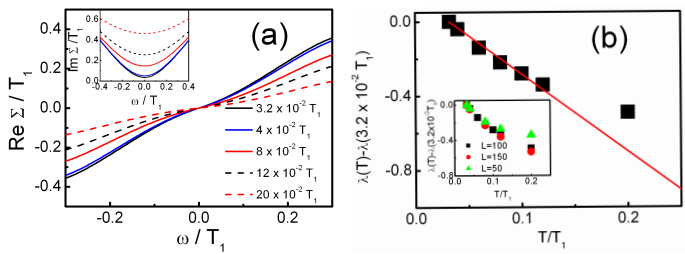,width=9.0cm}
\caption{(color online) (a)
The real part of the self-energy for various temperatures in the
normal state (from the numerical solution of
Eqs.(\protect\ref{n_1})-(\protect\ref{n_2} (\protect\ref{n_1}) and
Kramers-Kronig transform). The inset shows the imaginary part of the
self-energy. (b) The temperature dependence of the coupling constant
$\lambda (T) = d Re \Sigma (\omega \rightarrow 0, T)/d \omega$. The
straight line is the analytical result at low $T$, Eq.
\protect\ref{n_22}). The inset shows the $\lambda(T)$ for three
different $L$. } \label{fig1}
\end{figure}

To estimate these regular corrections, we computed the velocity
renormalization numerically. We used Eq. \ref{n_1} as a point of
departure, evaluated Im$\Sigma ({\bar k}, {\bar \omega})$ by
explicit 3D integration, and subtracted  antisymmetric contributions
to Im$\Sigma$ which do not affect linear in $\omega$ term in
Re$\Sigma$ (such contributions are generally present because the
dispersion is not particle-hole symmetric). In Fig. \ref{fig1}a we
present the results for the real part of the self-energy for various
temperatures, measured in units $T_1$. We see that the slope of the
real part of the self-energy decreases with increasing temperature
and is negative. In Fig. \ref{fig1}b we plot the $T$-dependent part
of $\lambda (T)$ vs Eq. (\ref{n_22}). We see that the agreement is
perfect up to approximately $0.1 T_1$. At larger $T$, the actual
$\lambda (T)$ flatters. In the insert to this figure, we show that
the $T$ dependent part of $\lambda$ (but not $\lambda (0)$) is
unsensitive to the upper cutoff of the momentum integration, in
agreement with Eq. (\ref{n_22}).
%

%AC shortened
The linear in $T$ dependence of the Fermi velocity cannot be carried
over to the $2k_F$ QC, non-Fermi liquid regime, because the
pre-factor for the $T$ term contains the divergent static spin
susceptibility at $2k_F$ (via ${\bar g} \propto K$). This
quantum-critical regime is relevant for electron-doped cuprates, in
which antiferromagnetism emerges near the electron density  at which
the Fermi surface passes through $(\pi/2,\pi/2)$,  i.e , along zone
diagonal $2k_F = (\pi,\pi)$ (Refs.\cite{ioffe,krotkov}).
%AC
We computed the velocity renormalization in the QC regime and found
$\omega = {\tilde \epsilon}_k (1 - 0.82 ({\tilde
\epsilon}_k/T)^{1/4})$, where ${\tilde \epsilon}_k \propto
(k-k_F)^{4/3}$. Observe  that a given $k$, $\omega$ still increases
with increasing $T$.

%The $2k_F$ QC behavior is quite involved and  fermionic self-energy
%crosses over from $\omega^{0.85}$ at the smallest frequencies, to
%$\omega^{3/4}$ at frequencies above some $\omega_0$, both are
%non-Fermi liquid forms~\cite{ioffe,krotkov}. For the $\omega^{3/4}$
%regime, the  regular, $q^2$ momentum dependence of the static $\chi
%(q-2k_F)$ is important. In the non-Fermi liquid regime, Fermi
%velocity is formally infinite, but one still can consider the
%temperature dependence of the quasiparticle dispersion, which tracks
%the position of the  maximum of the ARPES intensity. We verified
%that both quantum-critical regimes display $\omega/T$ scaling, such
%that at the lowest temperatures the correction to the
%zero-temperature dispersion scales as $T^{-0.15}$, while above
%$\omega_0$ it scales as $T^{-0.25}$. For the latter regime, we found
%explicitly that the dispersion is $\omega = {\tilde \epsilon}_k (1 -
%0.82 ({\tilde \epsilon}_k/T)^{1/4})$, where ${\tilde \epsilon}_k
%\propto (k-k_F)^{4/3}$. Observe  that a given $k$, $\omega$ still
%increases with increasing $T$.

To conclude, in this paper we considered temperature dependence of the nodal
Fermi velocity in 2D systems, $v_F (T) = v_F (0) (1 + \delta (T))$
We have found that $\delta (T)$ is positive and  linear in $T$ in
any Fermi liquid. The linear in $T$ term comes from the screening of
the interaction by just one particle-hole bubble. The corrections
due to extra dynamical screening only give rise to regular, $T^2$
terms. The slope of the linear term is quite large in the cuprates,
and agrees with the measurements on optimally doped $Bi2212$. In the
quantum-critical regime, the correction to the zero-temperature
dispersion is singular, and scales as $1/T^{0.15}$ at the lowest
$T$, and as $1/T^{0.25}$ at intermediate temperatures.

The experiments by Plumb et al~\cite{dess} have been performed only
at optimal doping. In the theory, the prefactor of the $T$ term
increases with decreasing doping, hence the prefactor should
decrease in the overdoped regime, and increase in the underdoped
regime. The measurements of the doping dependence of the slope are
clearly called for.

AVC acknowledges support from NSF-DMR 0604406  and from DFG via
Merkator GuestProfessorship, and is thankful to TU-Braunshweig for
the hospitality during the completion of this work. IE is supported
by the DAAD under Grant No. D/05/50420. We thank P.W.Anderson, D.
Dessau and M.R. Norman for useful discussions.


\begin{thebibliography}{99}
\bibitem{corr1} If the self-energy was analytic, $Im \Sigma (\omega)$
would be expandable in even powers of $\omega^2$, and $Re \Sigma$ in
odd powers of $\omega^2$. The $T$ dependence of the velocity then
would be $T^2$. Note also that in arbitrary $D <3$, $Im \Sigma
\propto \omega^2 + O(\omega^D)$, and the $T$ dependence of the
velocity is $T^{D-1}$.
\bibitem{comm_numb}
In earlier works~\protect\cite{acs,abanov}, one of us used a smaller
$U \sim 1eV$ to fit ARPES data near the hot spots.
 This was based on earlier estimate of $v_F$, which was roughly two times
smaller than $1eV$.  To get the same fits using $v_F =1eV$, one needs $U =2eV$.
\end{thebibliography}
\end{document}